\documentclass[%
 reprint,
superscriptaddress,
 amsmath,amssymb,
 aps,twocolumn,
pra,reprint,floatfix
]{revtex4-2}

\usepackage{graphicx}% Include figure files
\usepackage{dcolumn}% Align table columns on decimal point
\usepackage{bm}% bold math
\usepackage{dsfont}
\usepackage{blindtext}
\usepackage{mathtools}
\usepackage[utf8]{inputenc}
\usepackage[utf8]{inputenc}
\usepackage{booktabs}

\begin{document}

\author{Ahmed Gaber Abdelmagid}
\affiliation{Department of Mechanical and Materials Engineering, University of Turku, Turku, Finland}

\author{Hassan A. Qureshi}
\affiliation{Department of Mechanical and Materials Engineering, University of Turku, Turku, Finland}

\author{Michael A. Papachatzakis}
\affiliation{Department of Mechanical and Materials Engineering, University of Turku, Turku, Finland}

\author{Olli Siltanen}
\email{olmisi@utu.fi}
\affiliation{Department of Mechanical and Materials Engineering, University of Turku, Turku, Finland}

\author{Manish Kumar}
\affiliation{Department of Mechanical and Materials Engineering, University of Turku, Turku, Finland}

\author{Ajith Ashokan}
\affiliation{Chemistry Department, Clark Atlanta University, Atlanta, Georgia 30314, United States}

\author{Seyhan Salman}
\affiliation{Chemistry Department, Clark Atlanta University, Atlanta, Georgia 30314, United States}

\author{Kimmo Luoma}
\affiliation{Department of Physics and Astronomy, University of Turku, Turku, Finland}

\author{Konstantinos S. Daskalakis}
\email{konstantinos.daskalakis@utu.fi}
\affiliation{Department of Mechanical and Materials Engineering, University of Turku, Turku, Finland}

\title{Identifying the origin of delayed electroluminescence in a polariton organic light-emitting diode}

\date{September 2023}

\maketitle

\onecolumngrid

\section{Abstract}
Modifying the energy landscape of existing molecular emitters is an attractive challenge with favourable outcomes in chemistry and organic optoelectronic research. It has recently been explored through strong light-matter coupling studies where the organic emitters were placed in an optical cavity. Nonetheless, a debate revolves around whether the observed change in the material properties represents novel coupled system dynamics or the unmasking of pre-existing material properties induced by light-matter interactions. Here, for the first time, we examined the effect of strong coupling in polariton organic light-emitting diodes via time-resolved electroluminescence studies. We accompanied our experimental analysis with theoretical fits using a model of coupled rate equations accounting for all major mechanisms that can result in delayed electroluminescence in organic emitters. We found that in our devices the delayed electroluminescence was dominated by emission from trapped charges and this mechanism remained unmodified in the presence of strong coupling.

\section{Introduction}

Polariton chemistry has emerged as a promising new platform for modifying the molecular energy landscape, thus providing control over the photophysical and photochemical processes at room temperature~\cite{Sanvitto2016,Hertzog2019,Yuen-Zhou2020,Garcia-Vidal2021,Bhuyan2023}. Polaritons in planar optical microcavities are eigenstates resulting from strong coupling between the cavity modes and the molecular excited states in a material. In the simple picture of coupling one exciton resonance and one cavity mode, two eigenstates emerge which are called upper polariton (UP) and lower polariton (LP) with energies above and below that of the exciton resonance, respectively. The energy gap between UP and LP is called the vacuum Rabi energy, $\Omega$,  which scales up with increasing the number of active molecules, N, in the cavity mode volume, V, as$\sqrt{N/V}$. Experimentally, a pragmatically attractive property of optical microcavities is the ease with which one can tune the LP at a specific energy level by simply controlling the cavity thickness. This tunability presents an intriguing opportunity to explore the possibility of modifying the optoelectronic properties of molecular semiconductor materials and devices. 
Similarly to molecular design, by modifying the microcavity parameters the LP mode can be tuned to energies that match that of the triplet states~\cite{Lee2017}. This could potentially assist or even facilitate triplet-to-singlet population migration via mechanisms such as reverse intersystem crossing (RISC) or triplet-triplet annihilation (TTA). 

Under optical excitation, there are studies that investigate the effects of polaritons on RISC and TTA~\cite{Stranius2018, Martinez-Martinez2018, Berghuis2019, Eizner2019, Polak2020, Yu2021,Mukherjee2023}. Currently, the main debate is around the collective nature of strong coupling in organic films due to the highly delocalized photon content in the polariton mode, which dilutes the polariton effect in intramolecular nonradiative processes~\cite{Tichauer2021,Martinez-Martinez2019,Sanchez-Barquilla2022,Miwa2023}. This means that the dominant mechanism for populating the polariton modes is through the exciton reservoir, either by radiative pumping or vibrationally assisted scattering~\cite{Hulkko2021}, and possible direct RISC from the T$_1$ state to the LP will occur at a negligible rate.
In materials with pre-existing high rates of triplet-to-singlet population transfer, namely thermally activated delayed fluorescence (TADF) and TTA, one can expect that it is difficult to experimentally resolve a process occurring at a negligible rate, hindering its observation. To gain further understanding, it is beneficial to extend these studies to fluorescent emitters.

\begin{figure*}[t!]
 \centering
 \includegraphics[width=0.6\linewidth]{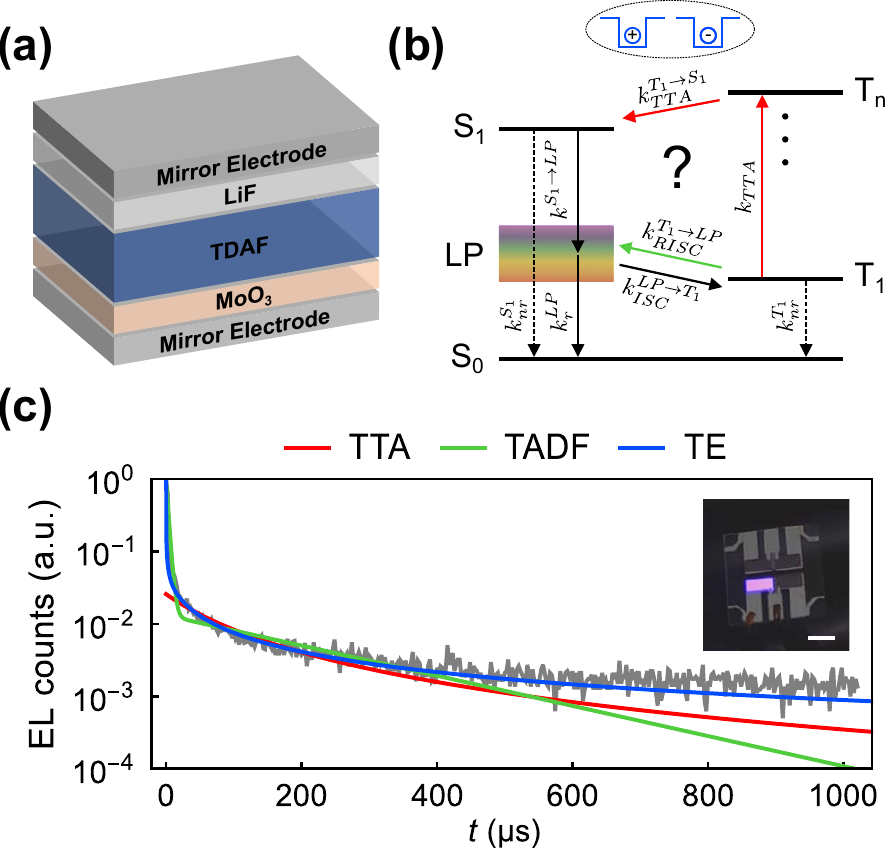}
 \vspace{0cm}
  \caption {
  Overview of the study. (a) Schematic representation of the bottom-emitting POLED structure consisting of an aluminium bottom electrode (30 nm), MoO$_3$ hole injection layer (5 nm), TDAF emitting layer, LiF electron injection layer (1 nm), and an aluminium top electrode (100 nm). For POLEDs with different LP resonants, we tuned the cavity resonance by adjusting the TDAF thickness. (b) Energy landscape for the used POLED with the possible relaxation pathways. (c) Transient electroluminescence of POLED~1 (grey) with the fitting results using the TTA model (red), TADF model (green), and TE model (blue). The inset in panel (c) shows a photograph of the POLED. The scale bar is 4~mm.
  }
  \label{fig:Fig1}
 \end{figure*}

Here for the first time, we study the time-resolved electroluminescence (EL) from bottom-emitting polariton organic light-emitting diodes (POLEDs) comprising a single fluorescent emitting layer of 2,7-Bis[9,9-di(4-methylphenyl)-fluoren-2-yl]-9,9-di(4-methylphenyl)fluorene (TDAF) sandwiched between aluminium mirror electrodes and injection layers of holes (MoO$_3$) and electrons (LiF). Figure\ref{fig:Fig1}(a) shows the schematic of the investigated devices. Under low injection current densities, delayed EL was recorded from the LP mode. We studied its origin and possible connection to the energy gap $\Delta$E$_{LP-T_1}$ as illustrated in Figure\ref{fig:Fig1}(b).

%\newpage
\section{Results}
\subsection{Steady-state measurements}
\begin{figure*}[t!]
 \centering
 \includegraphics[width=0.8\linewidth]{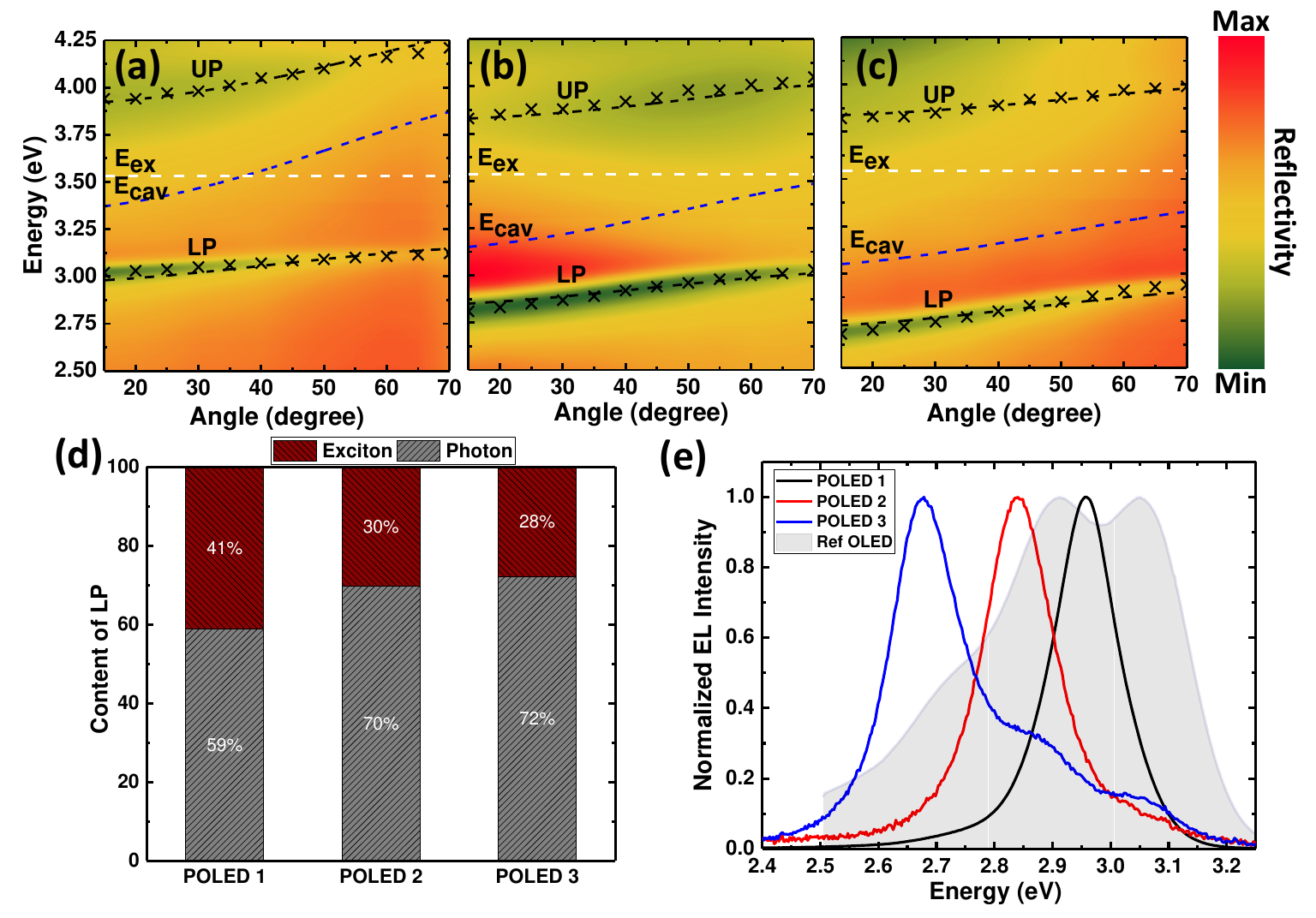}
 \vspace{0cm}
  \caption{
  Polariton characteristics. Angle-resolved reflectivity of (a) POLED 1, (b) POLED 2, and (c) POLED 3. The dashed white line is the molecular exciton energy, the dashed blue line is the cavity energy dispersion, the dashed black lines are fitted polariton dispersions, and the cross points are the experimental reflectivity minima. (d) Exciton (red berry) and photon (grey) content of the LP extracted from the coupled harmonic oscillator model at 15 degrees. (e) Normalized EL spectra of the different POLEDs and the reference device collected at a normal angle.}
 \label{fig:Fig2}
 \end{figure*}

TDAF is a well-established polaritonic organic semiconductor that has been used in studies of polariton lasing and superfluidity and exhibits a Rabi-splitting of $\sim$~1~eV~\cite{Kena-Cohen2013,Daskalakis2014,Lerario2017}. The latter is rather important as it enables us here to reach large $\Delta$E$_{S_1-LP}$
without compromising strong coupling (Figure~\ref{fig:Fig2}). Density functional theory (DFT) calculations performed at the screened range-separated hybrid LC-whPBE level and refined by measured S$_1$ and T$_1$ levels~\cite{TDAF_Triplet} reveal a large energy gap between S$_1$ and T$_1$ ($\Delta$E$_{S_1-T_1}$) of $\sim$~0.8~eV, which is essential in this study as it hinders RISC from T$_1$ to S$_1$ within the molecule. The landscape and character of the TDAF energy levels are further discussed in Supplementary Figure~\ref{DFT}. Moreover, TDAF's ambipolar electrical character makes it a favourable material in POLED studies~\cite{Gubbin2014,Daskalakis2019}, and here it allows us to directly populate the triplet states under electrical injection due to the spin-statistic rule (25 $\%$ singlets and 75 $\%$ triplets).

Figure~\ref{fig:Fig1}(c) shows a typical time-resolved EL measurement from our POLEDs along with a picture of a blue-emitting POLED in the inset. The POLED having an LP at 2.95~eV yields blue emission with a full-width at half-maximum of 0.13~eV and Commission Internationale de l’Eclairage (CIE) coordinates of (0.167, 0.015). In organic emitters, delayed EL from the S$_1$ level can be generally associated with either TTA~\cite{Wallikewitz2012}, TADF~\cite{Grune2020}, or slowly recombining charges in trapped states (trap emission---TE~\cite{Cheon2004}). This is shown in Figure~\ref{fig:Fig1}(c) together with the fittings from a rate-equation model that we present in Section~\ref{sec:rate-equation}. To investigate whether polaritons influence the dynamics of delayed EL in our POLEDs, we performed detuning- and injection current-dependent experiments. Our results demonstrate that the delayed EL mechanism remained the same regardless of the $\Delta$E$_{LP-T_1}$ or the existence of strong coupling. To identify the origin of the delayed EL, we carefully compared the experimental data with fittings from our model considering TTA, TADF, and TE parameters. Figure~\ref{fig:Fig1}(c) demonstrates that TE fitting is in perfect agreement with the experimental data.

We fabricated POLEDs with LP at 2.95~eV, 2.83~eV, and 2.67~eV. The corresponding reflectivity contour plots are shown in Figures~\ref{fig:Fig2}(a)--(c), respectively. Fittings of the coupled harmonic oscillator model to the reflectivity dip result in the Rabi-splittings of 0.92~eV, 0.88~eV, and 0.96~eV for POLED~1, POLED~2, and POLED~3, respectively, which agree with previous reports on TDAF in strong coupling~\cite{Kena-Cohen2013}. Using the same fitting, we also estimated the exciton and photon content in each POLED shown in Figure~\ref{fig:Fig2}(d). Interestingly, even in a very negatively detuned microcavity with LP at 2.67~eV, we find that the LP band bottom exhibits a large exciton content of 28~$\%$ and shows clear anticrossing (see Supplementary Figure~\ref{fig:Ref} for the individual reflectivity spectra). Note that the POLEDs used in reflectivity measurements were top-emitting to avoid absorption through the UV-absorbing MoO$_3$ layer, and the two POLED configurations showed identical delayed EL profiles albeit with some detuning shifts shown in Supplementary Figure~\ref{TopBottom}. 
Furthermore, the semitransparent aluminium mirror was thinned to 25~nm (instead of 30~nm) to have better visibility of the UP. We also fabricated TDAF organic light-emitting diodes (OLEDs) in which the bottom electrode was replaced by an indium tin oxide (ITO) transparent layer to eliminate the cavity mode, and we clarified that these reference devices did not exhibit strong coupling. We refer to this OLED device as the \textit{reference device} throughout this work. The angle-resolved reflectivity shown in Supplementary Figure~\ref{fig:reference}(b) has a Lambertian absorption response and the normal-angle EL [Supplementary Figure~\ref{fig:reference}(c)] is typical for uncoupled TDAF molecule emission, confirming that no polariton modes are supported in these devices. More information on the reference device is shown in Supplementary Figure~\ref{fig:reference}.

Figure~\ref{fig:Fig2}(e) shows the EL spectra of the studied bottom-emitting POLEDs at the normal collection angle. The POLED with emission at 2.95~eV shows a uniform Lorentzian distribution with a full-width at half-maximum of 0.13~eV, while the POLEDs with LP tuned at 2.83~eV and 2.67~eV have a full-width at half-maximum of 0.14~eV and 0.15~eV, respectively, and exhibit strong inhomogeneity. Comparing the POLEDs with the reference device's EL spectrum, shown as a greyed-out area in Figure~\ref{fig:Fig2}(e) and in Supplementary Figure~\ref{fig:reference}, we attribute this inhomogeneity to emission from the uncoupled excitons escaping through the 30-nm-thick aluminium mirror. It is worth noting that shifting the LP resonant closer to T$_1$ resulted in a substantial reduction of the EL intensity.  
To our advantage, the thickness variation for the selected detuning is $\sim$~10~nm, while TDAF is ambipolar and thus insensitive to small shifts of the carrier recombination zone ~\cite{Tao2011}.  
Previously, in TDAF polariton OLEDs, a hole-blocking layer (BPhen) was used between TDAF and LiF~\cite{Gubbin2014}, which was not used in our study because we found BPhen devices to degrade rapidly during our measurements.

\subsection{Time-resolved electroluminescence}

We excite our samples using square electrical pulses with rise and fall times of sub-9~ns and collect the time-resolved EL using a custom-built k-space and time-correlated single photon counting (TCSPC) spectroscopy setup. See the Supplementary Figure~\ref{fig:setup} for details of the experimental setup. To ensure the consistency of the time-resolved EL measurements, we control the excitation pulse duration and repetition rate to allow the system to reach a steady state before we turn off the electrical pulse and collect the emission statistics. The injected current density was controlled by increasing the excitation pulse voltage and measuring the current with an oscilloscope. The EL from the POLEDs was spatially and spectrally filtered before it was collected by the TCSPC sensor. To ensure the consistency and validity of our findings, all the measurements were performed using freshly made POLEDs that were kept in a vacuum of $\sim10^{-3}$~mbar. In addition, throughout the duration of the time-resolved EL measurement we were tracking that the total collected photon counts remained stable. In some cases, samples degraded due to exposure to ambient conditions or due to overuse, showing a quenched emission trend resulting in an inflation of their delayed EL. 
We discarded such results from our final evaluation. An example of this inflation due to sample damage is shown in Supplementary Figure~\ref{fig:TREL_degraded}.

To explore the effect of strong coupling in the delayed EL of TDAF OLEDs, we compared the POLED~1 and 
the reference device. 
As it is clearly shown in Supplementary Figure~\ref{fig:reference}(d) the delayed EL of the reference device is dominated by TE statistics, further proving that the EL mechanism in the TDAF remained unmodified by the presence of strong coupling. 

Figure~\ref{fig:Fig3}(a) shows the time-resolved EL from POLEDs~1--3 at an injection current of 90~mA/cm$^2$. Despite how closely we approached T$_1$ with the LP, we observed identical trends. Moreover, all POLEDs display identical matching trends for injection current densities varying from $\sim$~30~mA/cm$^2$
to $\sim$~150~mA/cm$^2$ (shown in Supplementary Figure~\ref{fig:TREL_POLED12}). This further confirms that the polariton-alignment effect in the delayed emission of TDAF, if any, is negligible and difficult to resolve in raw data. 
By increasing the current density, interestingly, we observed a small quenching in the delayed EL trends. Nevertheless, to identify its origin and current-induced quenching, we developed a rate equation model that was used to fit the experimental results. 
 
Spin-orbit coupling calculations (SOC) (see Supplementary Figure~\ref{DFT}) reveal that S$_1$-T$_2$ SOC is an order of magnitude larger than S$_1$-T$_1$ SOC. This indicates that under the right conditions, TDAF could demonstrate ``hot RISC''~\cite{Lin2021}. In our case, the LP mode of POLED~2 is aligned with $T_2$ and also possesses substantial excitonic content of 30~$\%$, thus acting potentially as a ``hot RISC'' channel directly populating LP from $T_2$ with a rate $k_{hRISC}^{T_2\rightarrow LP}>0$. As 
indicated by Figures \ref{fig:Fig3} and \ref{fig:Fig4}, we did not observe this. We speculate that such a scenario will have interesting implications for the device's performance, and it is perhaps interesting to investigate further in the future.

\subsection{Rate-equation model and fitting}
\label{sec:rate-equation}

The population dynamics in our system, following the pulse turn-off, can be approximated by the following system of coupled rate equations. Here, we account for the presence of TTA, TADF, and TE [cf. Figure~\ref{fig:Fig1}(b)] and consider both the strong \textit{and} weak coupling (i.e., reference device).
\begin{align}
    \frac{dS_1}{dt}&=\frac{1}{4}L-(k_r^{S_1}+k_{nr}^{S_1}+k^{S_1\rightarrow LP})S_1+k_{TTA}^{T_1\rightarrow S_1}T_1^2,
    \label{eq:singlet_density}\\
    \frac{dLP}{dt}&=-(k_r^{LP}+k_{ISC}^{LP\rightarrow T_1})LP+k^{S_1\rightarrow LP}S_1+k_{RISC}^{T_1\rightarrow LP}T_1,
    \label{eq:polariton_density}\\
    \frac{dT_1}{dt}&=\frac{3}{4}L-(k_{nr}^{T_1}+k_{RISC}^{T_1\rightarrow LP})T_1+k_{ISC}^{LP\rightarrow T_1}LP-k_{TTA}T_1^2.
    \label{eq:triplet_density}
\end{align}
Here, $S_1$, $LP$, and $T_1$ are the time-dependent populations of S$_1$, LP, and T$_1$. $L$ is the Langevin recombination rate describing trapped charges. We assume that the excitons formed by trapped charges obey the spin-statistic rule: 25 $\%$ populating S$_1$ (or exciton reservoir) and 75 $\%$ T$_1$. $k_{(n)r}^{S_1}$ is the (non)radiative rate of S$_1$, $k_r^{LP}$ is the radiative rate of LP, $k_{nr}^{T_1}$ is the nonradiative rate of T$_1$, $k^{S_1\rightarrow LP}$ is the rate of internal conversion from S$_1$ to LP, $k_{ISC}^{LP\rightarrow T_1}$ is the rate of intersystem crossing from LP to T$_1$, $k_{RISC}^{T_1\rightarrow LP}$ is the rate of reverse intersystem crossing from T$_1$ to LP, $k_{TTA}$ is the rate at which two first-order triplets annihilate, and $k_{TTA}^{T_1\rightarrow S_1}$ is the rate at which TTA populates S$_1$. Note that, in general, $k_{TTA}\neq k_{TTA}^{T_1\rightarrow S_1}$. In the strong-coupling regime, S$_1$ becomes the exciton reservoir and we have $k_r^{S_1}=0$, whereas all rates involving $LP$ vanish under weak coupling.

Substituting Eqs.~\eqref{eq:singlet_density} and~\eqref{eq:polariton_density} to the EL  intensity $I_{EL}\propto R:=k_r^{S_1}S_1+k_r^{LP}LP$, we get
\begin{equation}
I_{EL}\propto\frac{1}{4}L-k_{nr}^{S_1}S_1-\frac{dS_1}{dt}+k_{TTA}^{T_1\rightarrow S_1}T_1^2-k_{ISC}^{LP\rightarrow T_1}LP-\frac{dLP}{dt}+k_{RISC}^{T_1\rightarrow LP}T_1.
\label{eq:EL_int}
\end{equation}

Note that $I_{EL}$ consists of both the prompt and delayed part. Next, we solve the intensity of \textit{delayed} EL ($I_{DEL}$) in different scenarios. For reasons that we will discuss later, we normalize the solutions so that $I_{EL}(0)=1$.

\textbf{TTA scenario:} If TTA dominates, we have $\frac{dT_1}{dt}\approx-k_{TTA}T_1^2$. Solving for $T_1$, substituting to Eq.~\eqref{eq:EL_int} under a similar approximation, and normalizing, we arrive at (cf. Ref.~\cite{Wallikewitz2012})
\begin{equation}
I_{DEL}(t)\approx\frac{k_{TTA}^{T_1\rightarrow S_1}/R(0)}{\big[1/T_1(s)+k_{TTA}(t-s)\big]^2},\hspace{5pt}t\geq s\gg0.
\label{eq:TTA}
\end{equation}
Here, $s$ is some reference point of time belonging to the TTA-dominant regime.

\textbf{TADF scenario:} If TADF dominates and $k_{ISC}^{LP\rightarrow T_1}LP$ and $k_{RISC}^{T_1\rightarrow LP}T_1$ are of the same order of magnitude, we have
\begin{align}
    \frac{dLP}{dt}&\approx-k_{ISC}^{LP\rightarrow T_1}LP+k_{RISC}^{T_1\rightarrow LP}T_1,
    \label{eq:tadf_polariton}\\
    \frac{dT_1}{dt}&\approx-k_{RISC}^{T_1\rightarrow LP}T_1+k_{ISC}^{LP\rightarrow T_1}LP.
    \label{eq:tadf_triplet}
\end{align}
Again, we solve for $T_1$, substitute to Eq.~\eqref{eq:EL_int}, and normalize, this time obtaining (cf. Ref.~\cite{Grune2020})
\begin{equation}
I_{DEL}(t)\approx\big[(k_{ISC}^{LP\rightarrow T_1}+k_{RISC}^{T_1\rightarrow LP})T_1(0)-k_{ISC}^{LP\rightarrow T_1}\big(T_1(s)+LP(s)\big)\big]\exp\big[-(k_{ISC}^{LP\rightarrow T_1}+k_{RISC}^{T_1\rightarrow LP})t\big]/R(0),\hspace{5pt}t\geq s\gg0.
\label{eq:TADF}
\end{equation}

\textbf{TE scenario:} Finally, should TE dominate, we can see from Eq.~\eqref{eq:EL_int} that $I_{DEL}\propto\frac{1}{4}L$. Here, the Langevin recombination rate $L$ is defined as~\cite{Cheon2004}
\begin{equation}
L=\gamma\int_0^d\rho_e(x,t)\rho_h(x,t)dx,
\label{eq:Langevin}
\end{equation}
where $\gamma$ is the bimolecular rate constant and $\rho_{e(h)}(x,t)$ is the density of trapped electrons (holes). Assuming that the charges are normally distributed over the recombination zone of thickness $d$~\cite{Cheon2004}, i.e.,
\begin{equation}
\rho_{e(h)}(x,t)=\frac{N_{e(h)}}{\sqrt{4\pi D_{e(h)}t}}\exp\Bigg[-\frac{(x-d/2)^2}{4D_{e(h)}t}\Bigg],
\label{eq:Gauss}
\end{equation}
with $N$ and $D$ denoting the initial concentrations and diffusion coefficients, $L$ becomes
\begin{equation}
L=\frac{\gamma N_eN_h}{2\sqrt{\pi(D_e+D_h)t}}\text{erf}\Bigg(\sqrt{\frac{\tau}{4t}}\Bigg).
\label{eq:langevin_solution}
\end{equation}
Here, $\tau:=d^2(D_e+D_h)/(4D_eD_h)$ is the characteristic recombination time of electrons and holes. Now, we can write the normalized delayed EL intensity as
\begin{equation}
I_{DEL}(t)\approx\frac{\gamma N_eN_h}{8R(0)\sqrt{\pi(D_e+D_h)t}}\text{erf}\Bigg(\sqrt{\frac{\tau}{4t}}\Bigg),\hspace{5pt}t\gg0.
\label{eq:langevin_del}
\end{equation}

\begin{figure*}[t!]
 \centering
 \includegraphics[width=.70\linewidth]{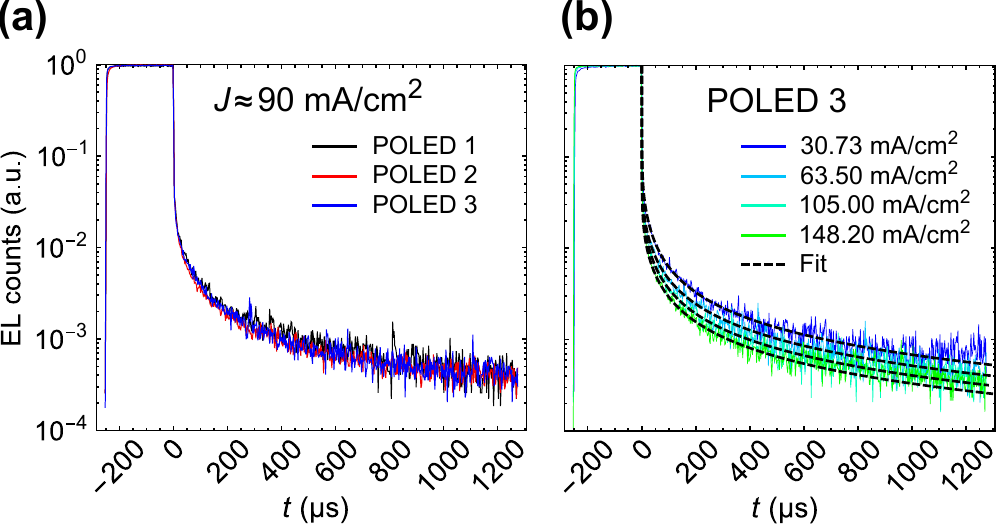}
  \caption {
  Time-resolved EL results and fittings. (a) Normalized EL counts of POLEDs 1--3 ($E_{LP}$ = 2.95~eV, 2.83~eV, 2.67~eV, respectively) at nearly the same current density. (b) Normalized EL counts of POLED 3 ($E_{LP}$ = 2.67~eV) and fitted TE functions (dashed curves) with four different current densities.
  }
 \label{fig:Fig3}
 \end{figure*}

\begin{table}
  \centering
  \caption{The mean absolute errors of the fittings.}
  \label{tbl:Table1}
  \begin{tabular}{ c c c c }
    \textbf{Device} & \textbf{TTA} & \textbf{TADF} & \textbf{TE} \\
    \midrule
    POLED 1 & 0.0006 & 0.0013 & 0.0004 \\
    POLED 2 & 0.0006 & 0.0011 & 0.0004 \\
    POLED 3 & 0.0004 & 0.0010 & 0.0002 \\
    Reference & 0.0009 & 0.0018 & 0.0006 \\
  \end{tabular}
\end{table}

Fitting Eqs.~\eqref{eq:TTA},~\eqref{eq:TADF}, and~\eqref{eq:langevin_del} to the time-resolved EL data, we find that the TE model fits the best [see Figures~\ref{fig:Fig1}(c), \ref{fig:Fig3}(b),  \ref{fig:reference}(d), and \ref{fig:residuals}]. The mean absolute errors calculated from all the fittings and the time span of 1~ms are given in Table~\ref{tbl:Table1}. From the errors, we see that also TTA could contribute to delayed EL. However, since the delayed EL in our case does not depend on detuning---as opposed to what might be expected with our energy landscape and TTA---we conclude that TE reigns over TTA, at least after the characteristic recombination time. With TADF, this is more apparent; Typically, TADF starts much earlier and its contribution dominates the overall EL intensity~\cite{Grune2020}. That is, we did not change the already negligible RISC rate of TDAF with strong coupling.

In Figure~\ref{fig:Fig3}(a), we have plotted the time-resolved EL data of POLEDs~1--3 ($E_{LP}=2.95\text{ eV, }2.83\text{ eV, }2.67\text{ eV}$) with nearly the same current density. We can clearly see that the delayed EL is independent of detuning.

Figure~\ref{fig:Fig3}(b) shows the time-resolved EL data of POLED 3 ($E_{LP}=2.67\text{ eV}$) and fit functions~\eqref{eq:langevin_del} with different current densities. The TE model describes our data extremely well---and although the model would seem to fit well with the prompt EL too, one should notice that $\lim_{t\to0}I_{DEL}(t)=\infty$. That is, prompt $I_{EL}$ near $t\approx0$ should be solved separately from $I_{DEL}$. In addition to the delayed EL models, we fitted monomials to the data and obtained approximately $1/t$-tails---a signature of trapped charges~\cite{Cheon2004}.

\begin{figure*}[t!]
 \centering
 \includegraphics[width=.75\linewidth]{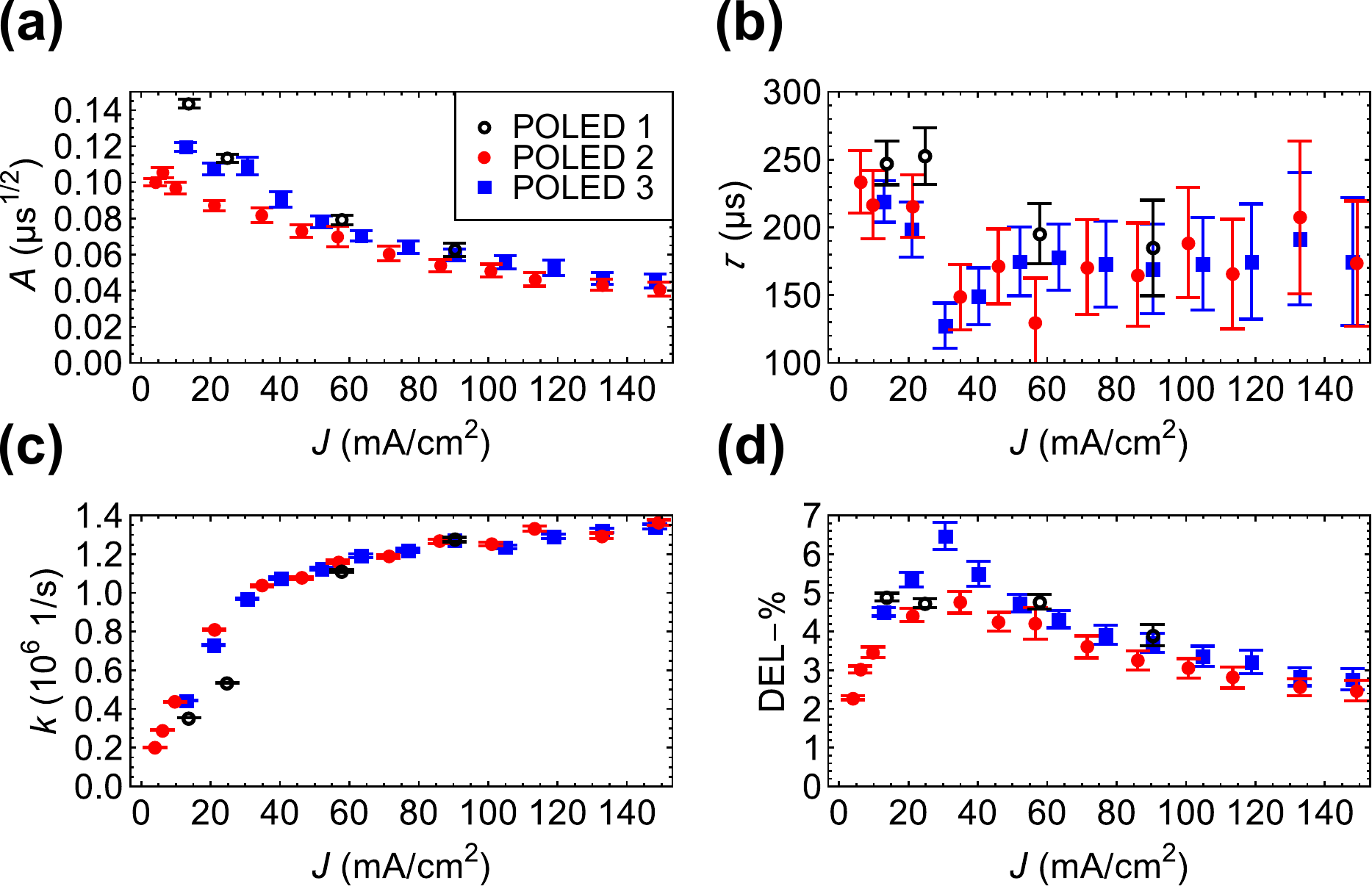}
  \caption{
  Parameters extracted from the fittings, as functions of current density. (a) The TE amplitude. (b) The characteristic recombination time. (c) The effective decay rate. (d) The DEL-\%. The error bars are standard deviations obtained from 100 independent fittings.
  }
 \label{fig:Fig4}
 \end{figure*}

It is of interest to evaluate the delayed emission contribution to the entire EL. We now define DEL-\% as the intersection of $I_{DEL}(t)$ and an exponential function $\exp(-kt)$ fitted on the prompt $I_{EL}(t)$ (cf.~\cite{Wallikewitz2012})---this is why we normalized the EL intensities. Here, $k$ is the effective decay rate of prompt EL. All the fitting results of POLEDs 1--3 are shown in Figure~\ref{fig:Fig4}. Figures~\ref{fig:Fig4}(a)--(d) show the TE amplitude $A:=\gamma N_eN_h/8R(0)\sqrt{\pi(D_e+D_h)}$, the characteristic recombination time $\tau$, the decay rate $k$, and the DEL-\%. Note that the resolution of prompt time-resolved EL may cause some error in our estimation procedure. Furthermore, the fitting of $\tau$ is quite sensitive to noise, which can explain the more fluctuating values in Figure~\ref{fig:Fig4}(b) when compared to other quantities. The error bars in Figure~\ref{fig:Fig4} were calculated using 100 perturbed data sets per current density and detuning. In each case, we simulated repeated measurements by adding white noise to the data, staying close to the original envelopes.

In Figure~\ref{fig:Fig4}(a), the TE amplitudes decrease smoothly---perhaps exponentially---while the other quantities behave more interestingly around $J_0\approx30\text{ mA/cm}^2$. Until this point, increasing current density means trapping more charges. Due to this aggregation, electrons and holes can recombine faster ($\tau$ becomes smaller), increasing the contribution of delayed EL. When we go beyond $J_0$, we start to promote different non-radiative processes such as singlet–singlet, singlet-triplet, and singlet-polaron annihilation~\cite{Murawski2013}, which dominate over the emission of trapped charges. That is, $\tau$ starts increasing and DEL-\% decreasing. Furthermore, as the singlets are involved in these processes, the effective decay rates in Figure~\ref{fig:Fig4}(c) increase at a slower rate.

\section{Discussion}
In conclusion, we studied the time-resolved EL of POLEDs. By comparing POLEDs and non-cavity OLEDs, we observed that delayed EL in our devices remained unchanged. Moreover, the LP modes were tuned within the energy landscape of the TDAF molecule. In particular, we explored the effect of matching LP to the T$_1$ energy level, while we scanned for changes in the dynamics. We concluded that this particular device system, despite offering a favourable test bed, did not show a prominent change in the dynamics either when introducing strong coupling or when moving the LP close to the T$_1$. These results suggest that strong coupling has a negligible effect on TDAF triplet management. In addition, we performed a comprehensive analysis of the time-resolved EL data using coupled rate equations that account for emission from the LP mode. Based on the analysis we performed in this particular experiment, we identified that the delayed EL in our devices originates from the recombination of charges trapped within the TDAF layer.

It is also worth noting that intermolecular near-neighbour processes---namely singlet fission, F\"{o}rster energy transfer, and TTA---are practically delocalized over many molecules and thus offer an attractive system to be influenced by photon-dressed polariton modes. Nevertheless, it is worth bringing up that in densely packed molecular films, such as the one usually implemented in microcavity polariton samples and OLEDs, intra- and intermolecular processes coexist and are often difficult to distinguish (e.g. RISC and TTA)~\cite{Mamada2023}.

Whether strong coupling can serve as the means for post-molecular design of materials with accelerated RISC and TTA is still an open question. There are enormous implications in such an approach as it could be used to address the low brightness (luminance) problem of OLEDs. This is a long-standing problem called efficiency roll-off. Macroscopically, it appears as a reduced internal quantum efficiency (IQE) at increased injection currents, while it microscopically originates from the microseconds-slow~\cite{Aizawa2022} delayed-fluorescence contribution to the EL IQE~\cite{Gather2011,mischok2023highly}. Importantly, strong coupling and photonics do offer an alternative route to investigate material properties that are usually inaccessible, and efforts towards this direction offer great future possibilities in the field of polariton chemistry.

\section{Methods}
\subsection{Fabrication}

The POLED devices were fabricated on pre-cleaned glass substrates using thermal evaporation at a base pressure below $10^{-7}$ Torr (Angstrom Engineering physical vapor deposition system). We used 15$\times$15 mm$^{2}$ glass substrates that were cleaned by sonication for 10 min in soapy water (3 $\%$ Decon 90), acetone, and isopropanol, respectively. The cleaned glass substrates were dried with nitrogen before device fabrication. A 30-nm-thick aluminium was deposited on top of the glass substrate as a bottom electrode, followed by deposition of 5~nm MoO$_{3}$ as the hole injection layer, TDAF as emitting layer, 1~nm LiF as the electron injection layer, and a 100-nm-thick aluminium as a top contact. The detuning of the POLEDs was controlled by varying the thickness of the emitting layer.    

\subsection{Characterization}
The angle-resolved reflectivity was measured with a J.A. Woollam VASE ellipsometer in reflectivity configuration. The EL was collected using a custom-made k-space setup (0.2 NA Microscope objective, 250~{\textmu m} slit width) consisting of a spectrometer coupled to a two-dimensional (2D) CCD camera (Princeton Instruments, 1340×400 pixels). Time-resolved EL was acquired using the same spectrometer and a pulse generator (HM8150) as an electrical excitation source. The POLEDs were excited electrically by 250~{\textmu s} pulse with different current densities. Further details can be found in Supplementary Figure \ref{fig:setup}.

%The sample was excited using 250 fs pulses at 350 nm and 200 kHz repetition rate (Light Conversion Pharos, Orpheus, and Lyra). 
\subsection{Computational Methodology}
The electronic structure calculations were performed by using the DFT at the screened range-separated hybrid (SRSH) method with optimally-tuned LC-$\omega$hPBE functional and 6-31G (d, p) basis set. The range separation parameter, $\omega$, was optimized using a minimization procedure based on the expression: J($\omega$) = [$\epsilon_{HOMO(\omega)}$ + IP($\omega$)]$^2$ + [$\epsilon_{LUMO(\omega)}$ + EA($\omega$)]$^2$. A dielectric constant of $\epsilon$~=~3.5 was considered for the SRSH calculations.  
The excited-state energies were estimated using the Tamm-Dancoff approximation (TDA) within the Time-dependent density functional theory (TDA-TDDFT) approach. The nature of the excited states was characterized using the Natural Transition Orbitals (NTO) analyses. The SOC values between the ground and excited states were estimated using the PySOC code interfaced with TDA-TDDFT calculations. These calculations are performed at two different dielectric constants, $\epsilon$~=~3.08 and 3.5, commonly used for such materials and following experimental conditions~\cite{Cooper2022,Sun2017}. Calculations by using two different dielectric constants reproduced similar trends. All DFT and TDA-TDDFT calculations were performed with the Gaussian16 program package~\cite{g16}.

\subsection*{Acknowledgements}
A.G.A. acknowledges the support from the Walter Ahlstr\"{o}m Foundation. The authors are grateful to David Lidzey and St\'{e}phane K\'{e}na-Cohen for their helpful comments on the results.

\subsection*{Funding}
This project has received funding from the European Research Council (ERC - http://dx.doi.org/10.13039/501100000781) under the European Union's Horizon 2020 research and innovation programme (grant agreement No. [948260]), and from Business Finland (http://dx.doi.org/10.13039/501100014438) project Turku-R2B-Bragg WOLED with decision number 1951/31/2021. The work at Clark Atlanta University was supported under the NSF Awards (1955299 and 2200387) and the ACCESS clusters (Bridges2 and Stampede-2) under Research Allocation Award TG-DMR200030 are acknowledged for computational resources and technical support.

\subsection*{Author Contributions}
K.S.D. conceived and guided the project. A.G.A. fabricated the samples. O.S. developed the theoretical model. O.S. and A.G.A. analyzed the experimental data. A.G.A., H.A.Q., and M.A.P. developed and performed the measurements. A.A. and S.S. performed the quantum chemical calculations. A.G.A., O.S., and K.S.D. wrote the manuscript. All authors contributed to the manuscript and analysis of the data and have accepted responsibility for the entire content of this manuscript and approved its submission.

\subsection*{Competing interests}
All authors declare that they have no competing interests.

%\bibliographystyle{plain}
%\bibliography{references}

\newpage
\setcounter{equation}{0}
\setcounter{figure}{0}
\setcounter{table}{0}
\makeatletter
\renewcommand{\theequation}{S\arabic{equation}}
\renewcommand{\thefigure}{S\arabic{figure}}

%\textbf{\LARGE \hl{A single emitter White-organic light-emitting diode}  }\\

\begin{center}
\textbf{\LARGE Supporting Information}\\
\end{center}
\begin{center}
\textbf{\LARGE{Identifying the origin of delayed electroluminescence in a polariton organic light-emitting diode}}\\
\end{center}
\noindent
Ahmed Gaber Abdelmagid, Hassan A. Qureshi, Michael A. Papachatzakis, Olli Siltanen, Manish Kumar, Ajith Ashokan, Seyhan Salman, Kimmo Luoma, and Konstantinos S. Daskalakis\\
\noindent
Corresponding authors: olmisi@utu.fi (O. Siltanen), konstantinos.daskalakis@utu.fi (K. S. Daskalakis)\\
\section*{Contents}

\begin{figure*}[h!]
 \centering
 \includegraphics[width=1\linewidth]{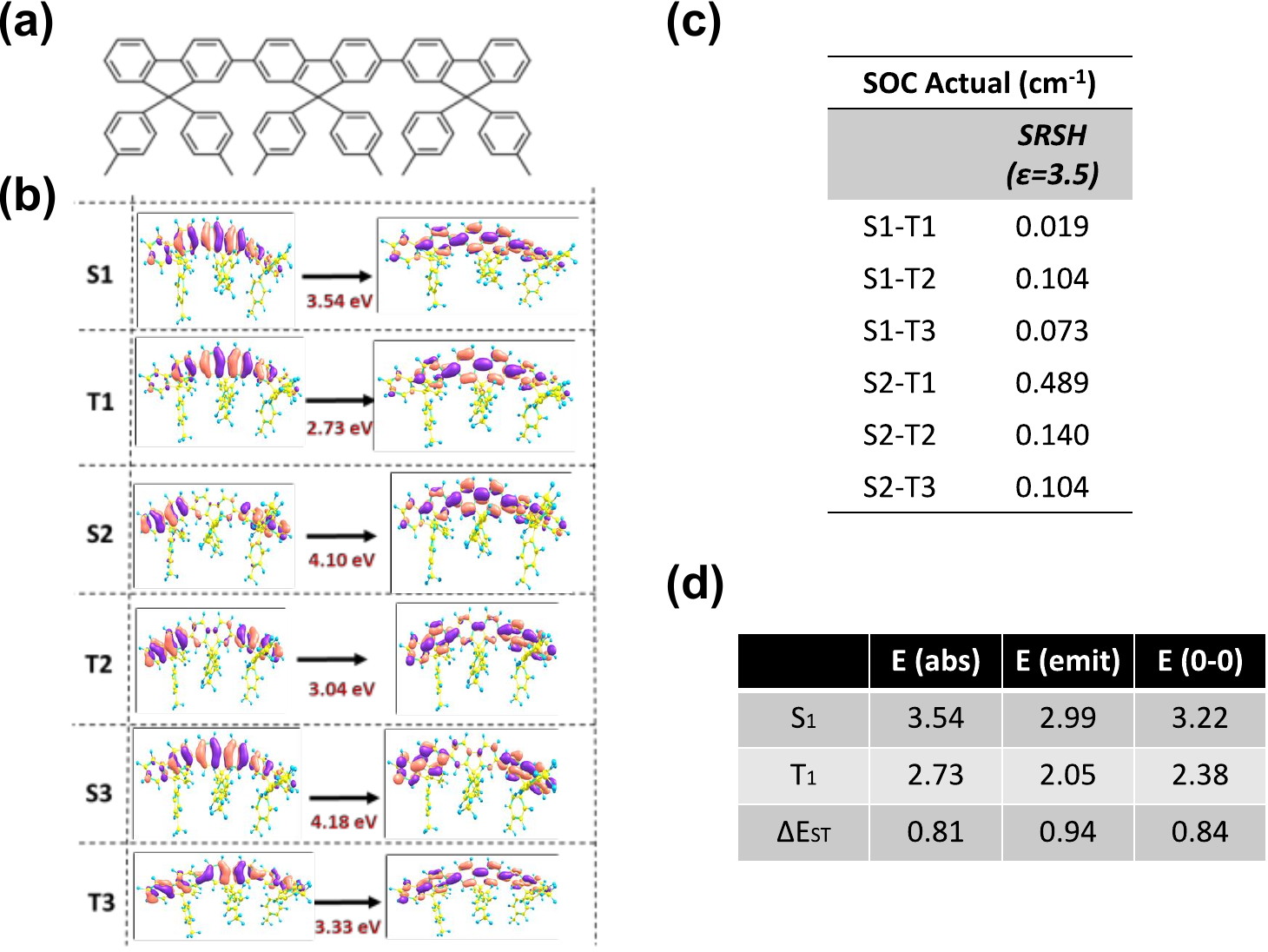}
 \vspace{0cm}
  \caption {(a) The molecular structure of TDAF. (b) Natural transition orbitals (NTOs) offer a representation of the wavefunctions of the hole and electron within a specific excited state, in the first three excited levels of the singlet and triplet states. The overlap between the hole and electron NTOs results in S$_1$ exhibiting characteristics of local excitation (LE), which implies the strong transition of the TDAF exciton. (c) Calculated spin-orbit coupling matrix elements. (d) Calculated vertical (abs), adiabatic (emit), and potential surface minimum (0-0) energies for the first singlet and triplet excited states in eV. These calculations were performed at Screened Range Separated Hybrid functional LC-whPBE with an implicit dielectric constant of 3.5. We opt for this approach as it offers the best alignment between our calculations, the earlier experimental findings of TDAF~\cite{TDAF_Triplet}, and our results. }
 \label{DFT}
 \end{figure*}

\begin{figure*}[h!]
 \centering
 \includegraphics[width=1\linewidth]{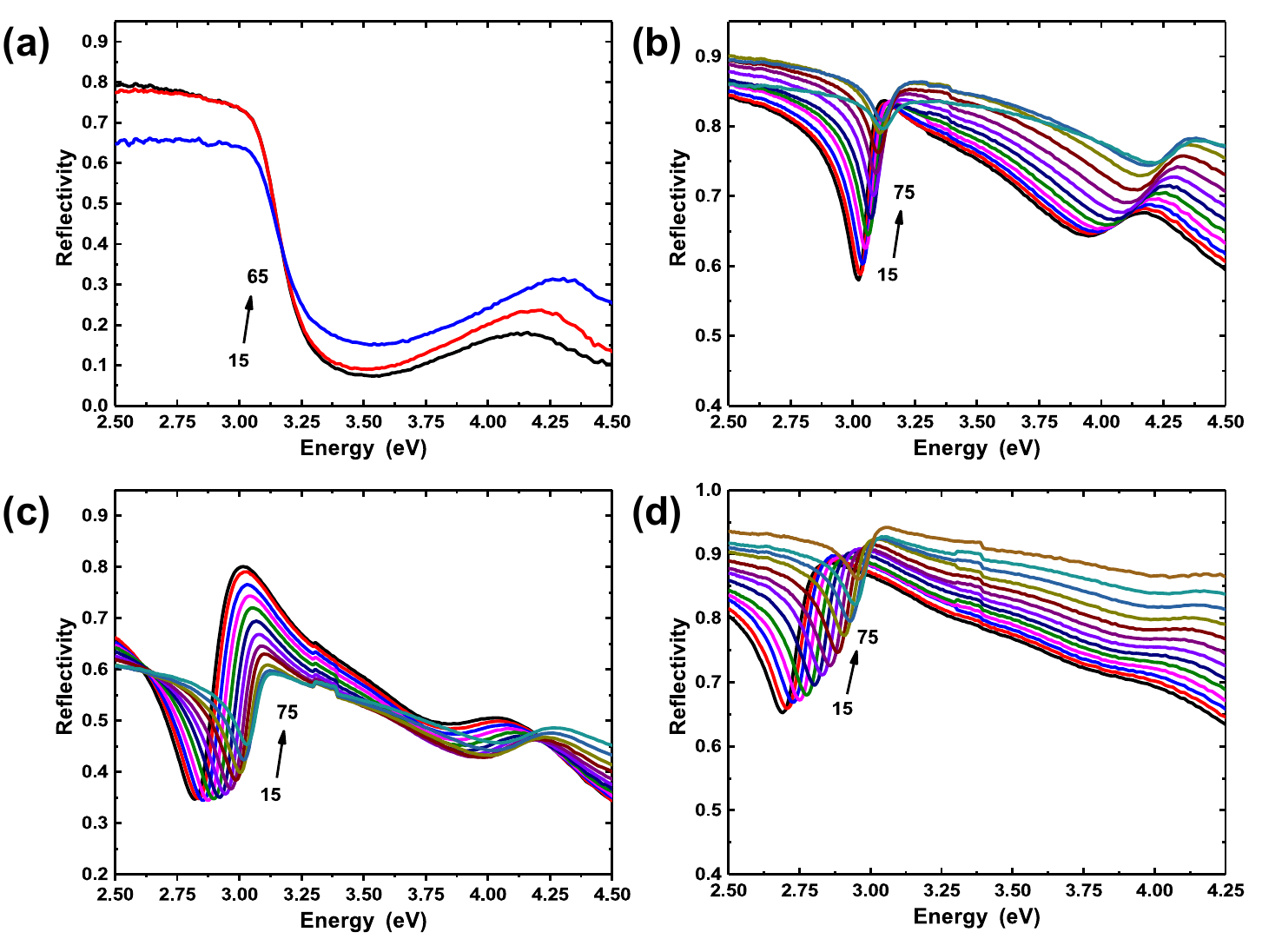}
 \vspace{0cm}
  \caption {Angle-resolved reflectivity for (a) ITO-based device with the angle interval 25$^\circ$---the data shows an angle independent spectrum which is similar to the emission of the TDAF molecule, i.e., no strong coupling effect---(b) POLED 1 with the angle interval 5$^\circ$, (c) POLED 2 with the angle interval 5$^\circ$, and (d) POLED 3 with the angle interval 5$^\circ$.}
 \label{fig:Ref}
 \end{figure*}

\begin{figure*}[h!]
 \centering
 \includegraphics[width=0.8\linewidth]{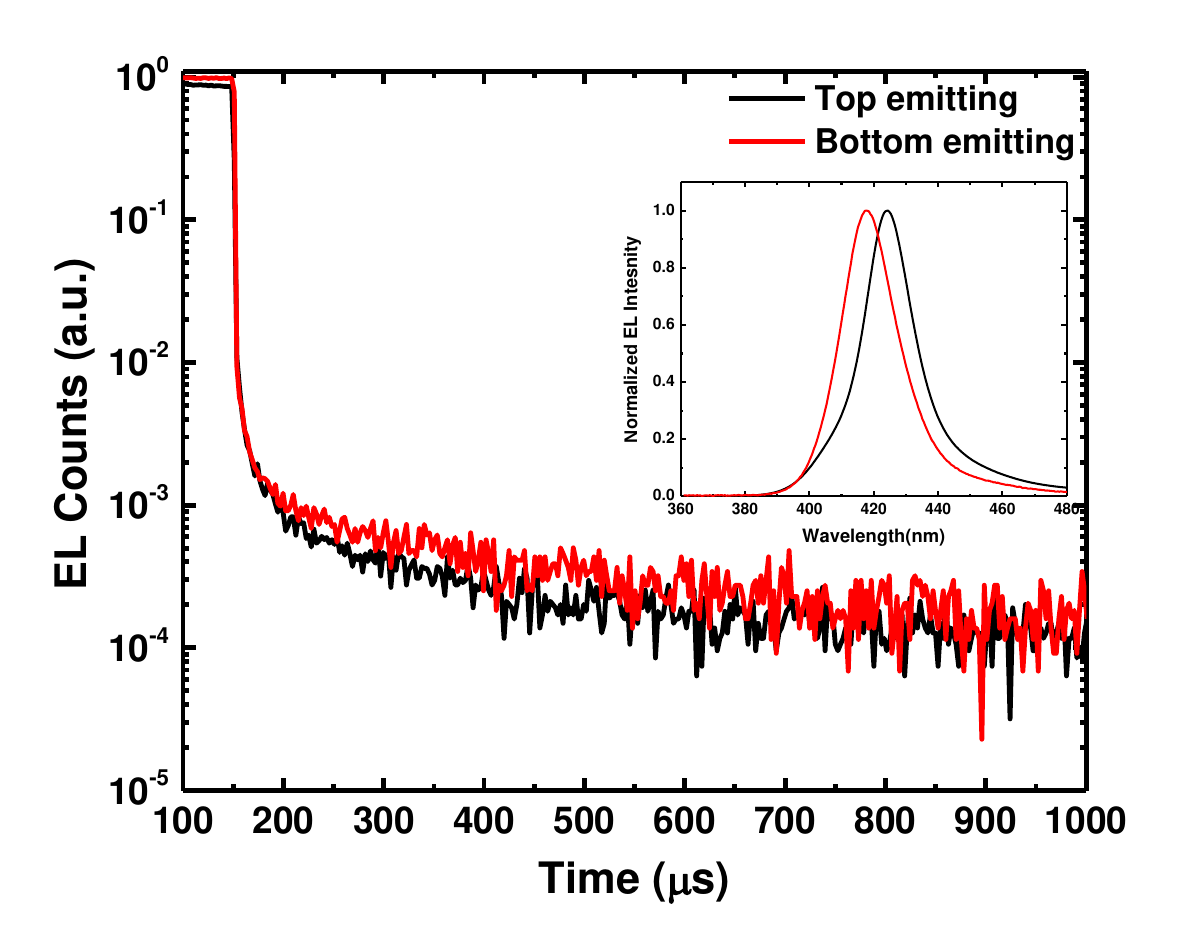}
 \vspace{0cm}
  \caption {Normalized time-resolved EL counts for a top-emitting POLED (black) consisting of an aluminium bottom contact (70~nm), MoO$_3$ hole injection layer (5~nm), TDAF emitting layer, LiF electron injection layer (1~nm), and an aluminium top contact (25~nm) and a bottom-emitting POLED (red) consisting of an aluminium bottom contact (30~nm), MoO$_3$ hole injection layer (5~nm), TDAF emitting layer, LiF electron injection layer (1~nm), and an aluminium top contact (100~nm). The inset shows the normalized EL spectra of both POLEDs. The slight difference in the delayed part of the EL is due to the difference in the top and bottom contact thicknesses of the two POLEDs.}
 \label{TopBottom}
 \end{figure*}

\begin{figure*}[h!]
 \centering
 \includegraphics[width=1\linewidth]{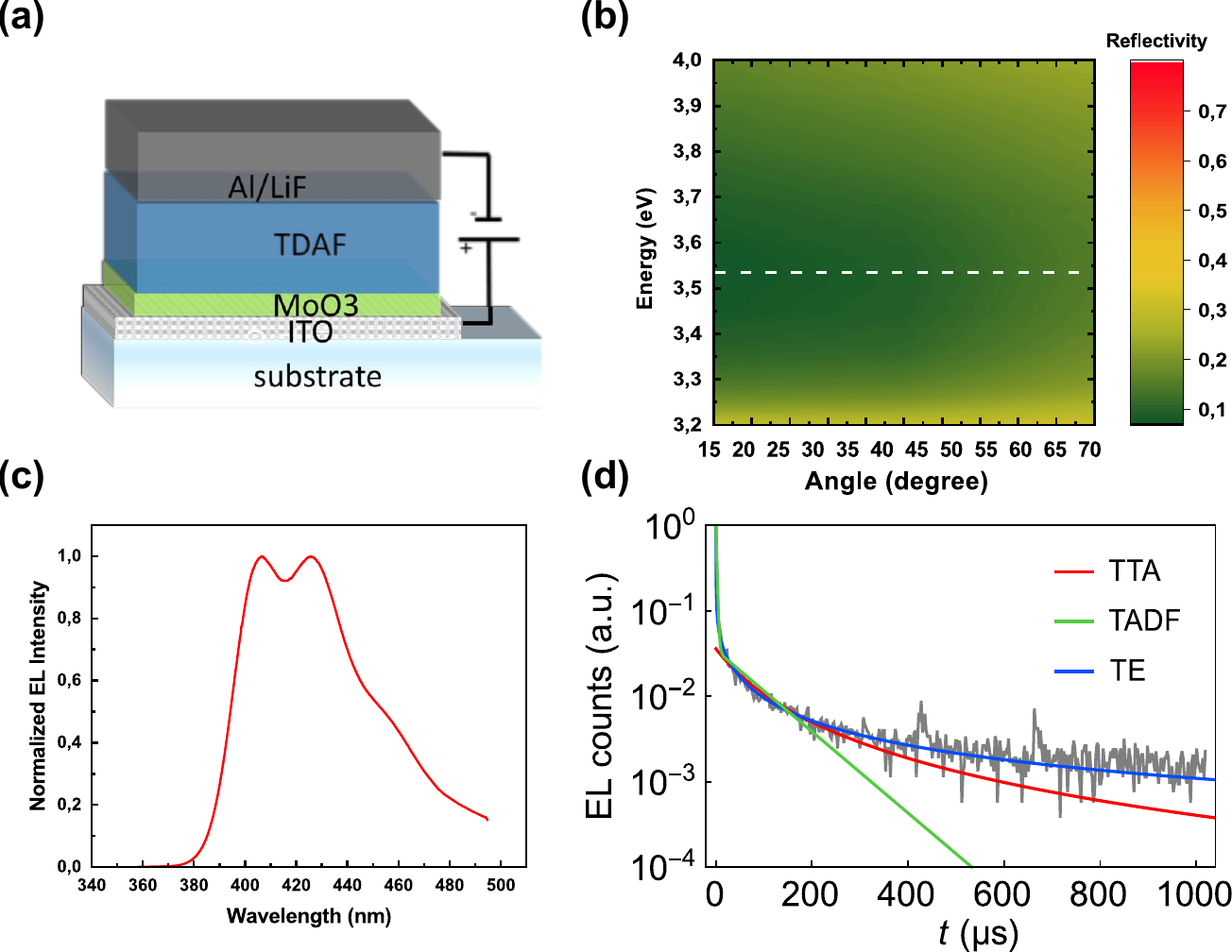}
 \vspace{0cm}
  \caption {(a) Schematic representation of the reference device. (b) Angle-resolved reflectivity of the reference device. (c) Normalized EL spectrum of the reference device. (d) Normalized time-resolved EL counts of the reference device at $J=75$ mA/cm$^2$ (grey) and the fitted models. Here, the RISC related to TADF happens from T$_1$ to S$_1$ (not LP). As expected, the model fits poorly.}
 \label{fig:reference}
 \end{figure*}

\begin{figure*}[h!]
 \centering
 \includegraphics[width=1\linewidth]{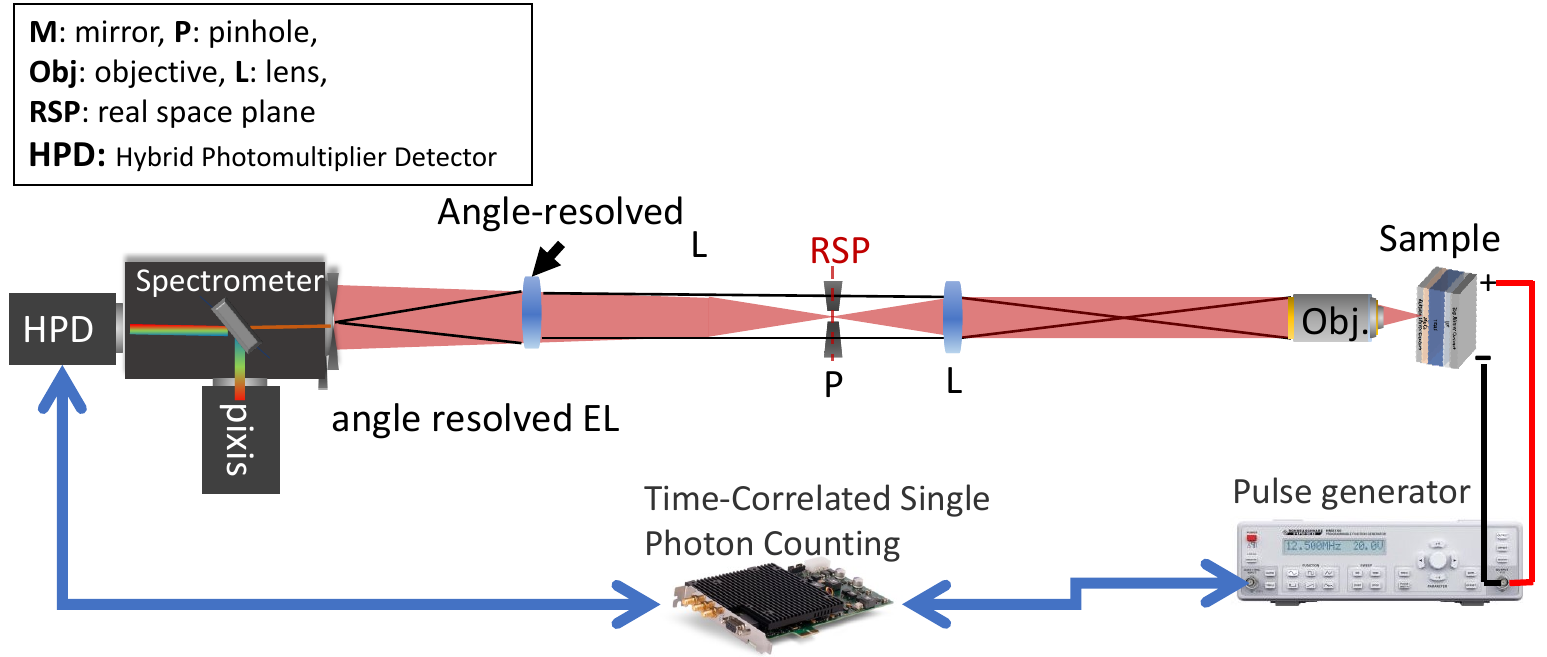}
 \vspace{0cm}
\caption{Schematic representation of the measurement setup. The electroluminescence (EL) was collected with a Nikon CFI Plan Apochromat Lambda 4X magnification and 0.2 NA. A 200 mm tube lens was used to focus the EL image of the samples on an iris to spatially filter out emission from the sides. The filtered image was transferred by a 200 mm relay lens to the slit of the spectrometer (Acton 300 Princeton Instruments) which has two outputs that can be switched via a mirror in the spectrometer. To verify the EL spectrum, the EL was directed to a 2D CCD (PIXIS:400B Princeton Instrument). To perform the TCSPC measurement, EL from the LP band-bottom was directed to the hybrid photomultiplier detector (PMA Hybrid PicoQuant) output. Current injection to the samples was realized by a pulse generator (HM8150 Rohde \& Schwarz) with 250~\textmu s square pulses with 9~ns turn-on time. We used a TimeHarp 260 TCSPC board module (PicoQuant) to track the photons and synchronize the pulse generator.}
 \label{fig:setup}
 \end{figure*}

\begin{figure*}[h!]
 \centering
 \includegraphics[width=1\linewidth]{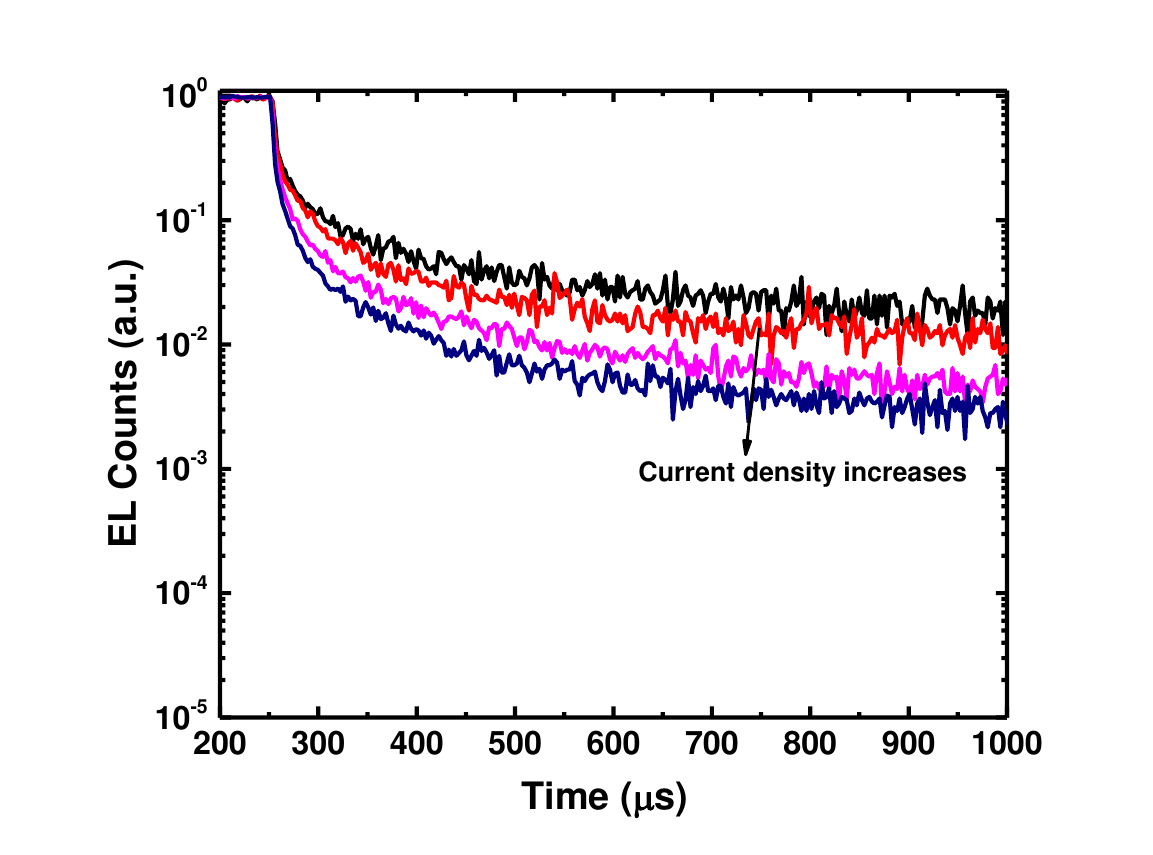}
 \vspace{0cm}
  \caption {Normalized time-resolved EL counts for a degraded POLED. The TE has increased as a result of the sample's deterioration subsequent to its exposure to ambient air over a period of time.}
 \label{fig:TREL_degraded}
 \end{figure*}

\begin{figure*}[h!]
 \centering
 \includegraphics[width=1\linewidth]{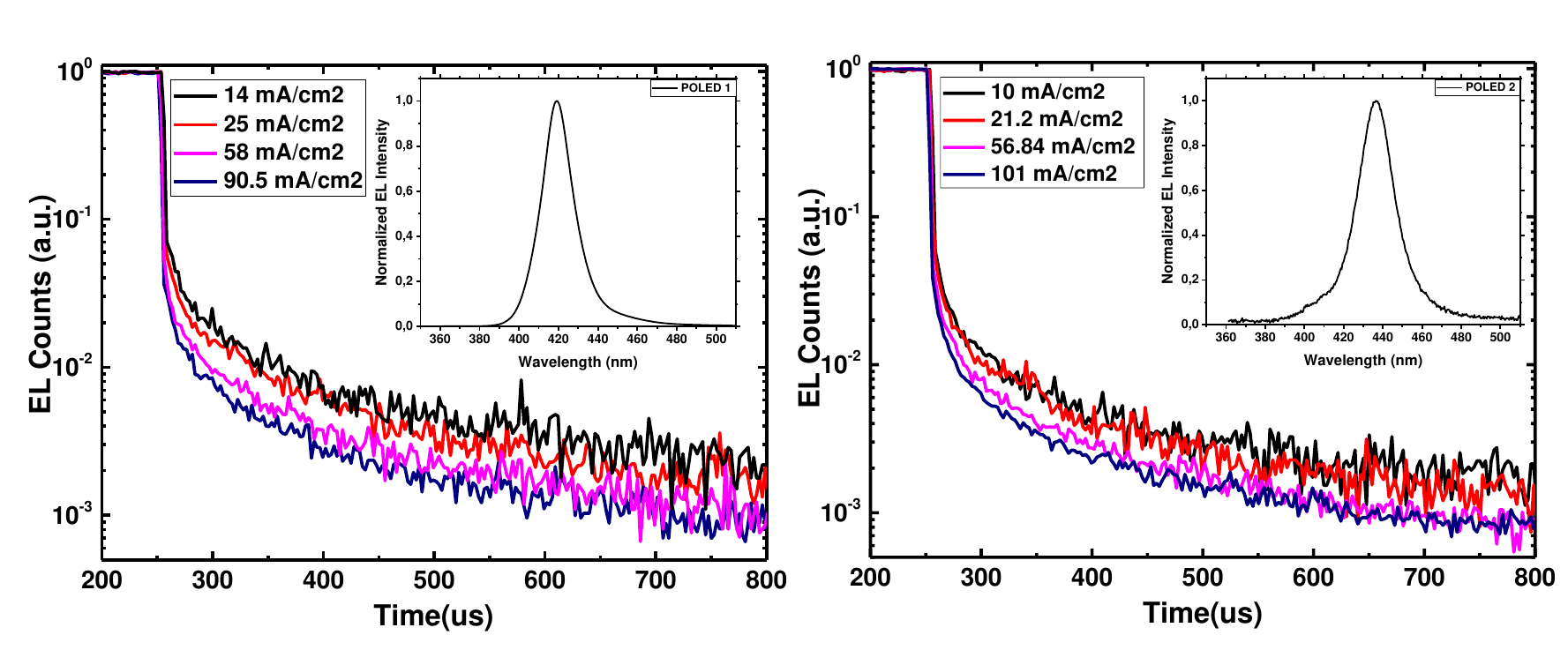}
 \vspace{0cm}
  \caption {Normalized time-resolved EL from (a) POLED 1 and (b) POLED 2. The insets are the normalized EL spectra. }
 \label{fig:TREL_POLED12}
 \end{figure*}

\begin{figure*}[h!]
 \centering
 \includegraphics[width=.60\linewidth]{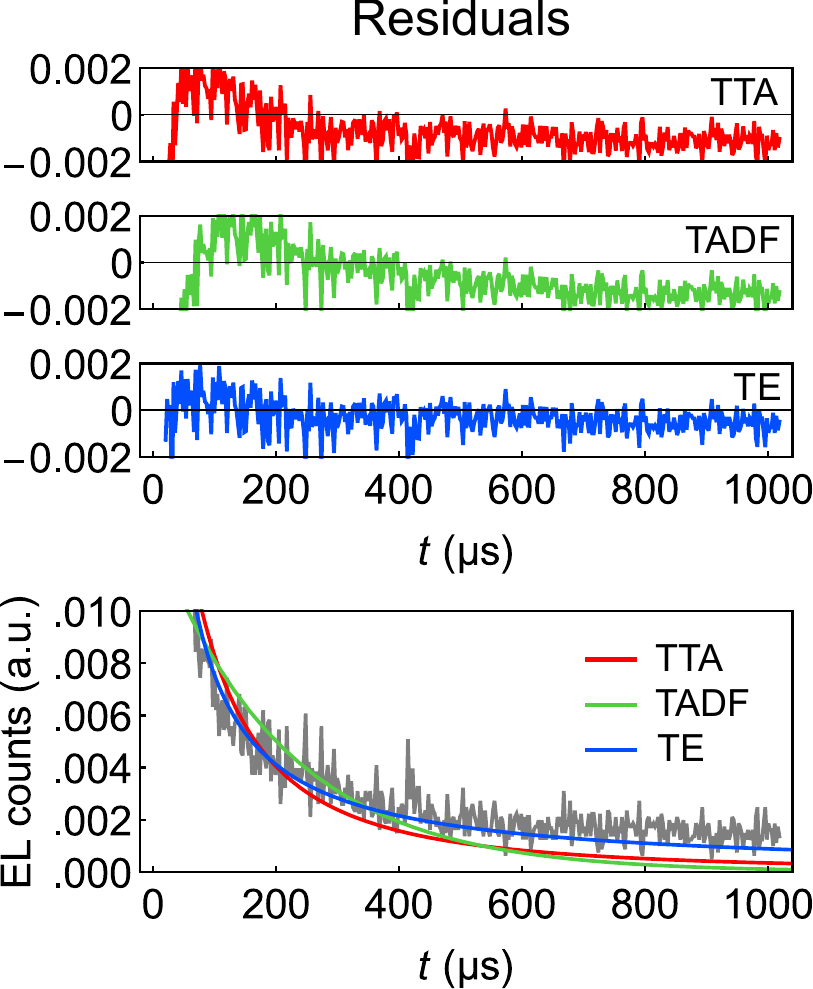}
 \vspace{0cm}
  \caption {Residuals of the different models of delayed EL with POLED 2 and $J=6.31$ mA/cm$^2$.}
 \label{fig:residuals}
 \end{figure*}

\noindent
\textbf{Supplementary Figure \ref{DFT}}. Density functional theory simulations.\\
\newline
\textbf{Supplementary Figure \ref{fig:Ref}}. Angle-resolved reflectivity of the reference device and POLEDs 1--3.\\
\newline
\textbf{Supplementary Figure \ref{TopBottom}}. Comparison between top- and bottom-emitting POLED.\\
\newline
\textbf{Supplementary Figure \ref{fig:reference}}. Steady-state and time-resolved measurements of the reference device.\\
\newline
\textbf{Supplementary Figure \ref{fig:setup}}. Measurement setup.\\
\newline
\textbf{Supplementary Figure \ref{fig:TREL_degraded}}. Degradation effect in time-resolved electroluminescence.\\
\newline
\textbf{Supplementary Figure \ref{fig:TREL_POLED12}}. Current-dependent time-resolved electroluminescence of POLED 1 and 2.\\
\newline
\textbf{Supplementary Figure \ref{fig:residuals}}. Fitting residuals of the different mechanisms on POLED 2.\\

\newpage  

\newpage 

\newpage 

\newpage

\newpage

\end{document}